\documentclass{statsoc}
\usepackage[a4paper]{geometry}
\usepackage{soul}
\usepackage{graphicx}
\usepackage[textwidth=8em,textsize=small]{todonotes}
\usepackage{amsmath}
\usepackage{natbib}
\usepackage{comment}
 \usepackage[normalem]{ulem}
 \useunder{\uline}{\ul}{}

\newcommand{\rthree}[1]{\textcolor{black}{#1}}

\title[Forks over knives]{Forks Over Knives: Predictive Inconsistency in Criminal Justice Algorithmic Risk Assessment Tools}
\author[Greene {\it et al.}]{Travis Greene}
\address{Institute of Service Science, National Tsing Hua University, Hsinchu, Taiwan}
\email{travis.greene@iss.nthu.edu.tw}
\author{Galit Shmueli}
\address{Institute of Service Science, National Tsing Hua University, Hsinchu, Taiwan}
\author{Jan Fell}
\address{Institute of Service Science, National Tsing Hua University, Hsinchu, Taiwan}
\author{Ching-Fu Lin}
\address{Inst. of Law for Science and Technology, National Tsing Hua University, Hsinchu, Taiwan}
\author[Greene et al.]{Han-Wei Liu}
\address{Department of Business Law and Taxation, Monash University, Australia}

\begin{document}
\enlargethispage{2pc}
\begin{abstract}
Big data and algorithmic risk prediction tools promise to improve criminal justice systems by reducing human biases and inconsistencies in decision making. Yet different, equally-justifiable choices when developing, testing, and deploying these sociotechnical tools can lead to disparate predicted risk scores for the same individual. Synthesizing diverse perspectives from machine learning, statistics, sociology, criminology, law, philosophy and economics, we conceptualize this phenomenon as \emph{predictive inconsistency}. We describe sources of predictive inconsistency at different stages of algorithmic risk assessment tool development and deployment and consider how future technological developments may amplify predictive inconsistency. \rthree{We argue, however, that in a diverse and pluralistic society we should not expect to completely eliminate predictive inconsistency. Instead, to bolster the legal, political, and scientific legitimacy of algorithmic risk prediction tools, we propose identifying and documenting relevant and reasonable ``forking paths" to enable quantifiable, reproducible multiverse and specification curve analyses of predictive inconsistency at the individual level.}
\end{abstract}

\section{Introduction: The Evolution of Algorithmic Risk Assessment Instruments}

Prediction and classification methods have played a role in criminal justice for over a century. For nearly as long, doubts regarding the suitability and performance of such methods have loomed. In 1895, for instance, Francis Galton wondered whether different judges might impose different patterns of penalties for the same kinds of offenders \citep{gottfredson1987prediction}. What started as “behavioral forecasts” informing parole decisions in the 1920s \citep{berkbleich2014forecasts} eventually evolved into more sophisticated “actuarial” prediction methods used since at least the 1970s in the United States (US). Since then, data-driven statistical and algorithmic risk assessment instruments (ARAIs) have been used in a variety of criminal justice contexts including sentencing, pretrial incarceration, parole, probation supervision levels, and security levels decisions \citep{brennan1987classification}. Machine learning (ML) algorithms are now deployed at numerous stages in the judicial process, from pre-guilt fact-finding to post-guilt sentencing stages, as part of a general trend towards “evidence-based sentencing” \citep{mckay2020predicting}. Although the acronym ARAI originally referred to “actuarial risk assessment instruments,” it now also refers to more sophisticated “algorithmic” tools \citep{partnershipAI2019}.

In an era of budget cuts and limited resources, US-based judges have increasingly become “managerial,” relying on the power of information systems to quickly resolve disputes and organize growing caseloads \citep{resnik1982managerial}. Consider the CompStat (computerized statistics) system used by police departments around the world since the 1990s \citep{bratton2008police}. CompStat and other tools like it combine data-driven decision making with private sector ``performance management'' techniques emphasizing standardized metrics to track progress and promote accountability. But lunches are not free. The growing pressures of bureaucratic efficiency and consistency—resulting in determinate sentencing laws, and, \rthree{in some jurisdictions, the mandatory use of risk assessment tools \citep{stevenson2018assessing}—have left judges with varying degrees of in overriding the tool's recommendation \citep{garrett2020judging}.} \rthree{Still, the} standardized pre-sentence reports compiled by probation officers have been criticized by some as a “ritual” used to “maintain the myth of individualized justice” \citep{rosecrance1988maintaining}. More recently, the 2016 \emph{State v. Loomis} case re-ignited similar debates when Loomis unsuccessfully argued in the Wisconsin Supreme Court that including a prediction from the COMPAS ARAI in his pre-sentencing investigation report violated his rights to due process and individualized sentencing \citep{liu2019beyond}.

Despite these concerns, the digitization of court records, new forms of data collection, and cheaper processing power have catalyzed the use of data analytics and algorithms in the decision-making processes of law enforcement agencies, corrections officials, and judges \citep{coglianese2021ai}. Today, over 200 ARAIs are used worldwide, and US corrections agencies alone rely on over 19 different tools \citep{duwerocque2017effects, fazel2018selecting}. Proponents of modern data-driven approaches to risk assessment emphasize their cost-effectiveness in prioritizing limited government resources and in predicting and controlling individual behavior. ARAIs promise to reduce human bias and inconsistency, and provide a scientific and evidence-based approach to the judicial process \citep{zavrvsnik2020criminal}. Data-driven risk assessment helps manage the complexity of judicial decision-making, which often involves estimating the likelihood of future unlawful behavior, given an individual’s unique personal characteristics, social connections, and past history \citep{monahan2016risk}.  In US pre-trial contexts, risk assessment tools are part of reformatory efforts aimed at reducing wealth-based disparities and providing “equal justice for poor and rich, weak and powerful,” in line with constitutional guarantees to equal protection and due process \citep{mayson2017dangerous}. The overall goal is to move the US towards a “smarter” regime by incorporating statistical and ML risk prediction algorithms into the criminal justice system \citep{berkhyatt2015machine}. Although the US is unique in its reliance on and early adoption of ARAIs, many countries are now implementing predictive algorithms in their criminal justice systems, including China \citep{lixiaohui2020research} and the United Kingdom (UK) \citep{cui2020artificial}. Others, such as Austria, Latvia, and the Netherlands, are debating their use \citep{councilEU2019AI}.

Experts fall into several camps about the appropriateness of ARAIs in criminal justice decision-making contexts. Some scholars view the algorithmization of judicial decision making as inevitable and argue courts should embrace automation \citep{volokh2018chief}. Well-designed (even if imperfect) algorithms can mitigate human judges’ biases and inconsistencies \citep{zavrvsnik2020criminal}, improve the criminal justice system \citep{corbett2017algorithmic}, and reduce prison and jail populations. ARAIs can help limit unwanted judicial discretion in sentencing by giving offenders with similar characteristics similar sentences, resulting in more consistent and rational sentencing policy \citep{frase2000guided}.  As \citet{eckhouse2019layers} explain, “With the [ARAI] scores as guidance, judges and policymakers can apply the same model to every case and claim they have used an objective, neutral mechanism of fair treatment.” 

Yet critics allege ARAIs are simply a new mode of “penological control” \citep{feeley1992new} adapted from managerial methods of risk assessment. Instead of rehabilitating, reintegrating, or retraining offenders, these methods merely shuffle the allocation of “risky” offenders in society by “selective incapacitation”  \citep{auerhahn1999selective}. And many scholars, particularly from science and technology studies, are skeptical of claims that the rationality and objectivity of big data justifies the use of ARAIs in criminal justice contexts \citep{boyd2012critical, moses2014using, dressel2018accuracy}. Complex and proprietary algorithms may inadvertently reflect and exacerbate existing social biases and discrimination embedded in training data \citep{liu2019beyond}, thus reproducing injustices already present in society \citep{mittelstadt2016ethics}. A recent and related stream of research draws from surveillance studies, critical race theory, sociology and feminist theory to argue that ARAIs express a form of “technoscientific” power whose ahistorical and ostensibly neutral algorithms, data, and computer code serve to maintain structural inequalities across racial and social boundaries \citep{eubanks2018automating, benjamin2019race, dignazio2020datafem}. 

\subsection{The Role of Prediction in Punishment and Sentencing}
Whether algorithmic risk assessment is considered a boon or bane may depend on one's theory of criminal punishment. Modern ARAI proponents often assume a utilitarian or consequentialist ethics in which public safety and deterrence is the primary goal (see, e.g., \citet{corbett2017algorithmic}). Sentencing and punishment thus assume a forward-looking perspective with a focus on minimizing the costs and probability of future unlawful behavior \citep{hamilton2015risk}. Indeed, some US states, such as Pennsylvania, require judges to consider defendants’ future dangerousness during sentencing \citep{berkbleich2014forecasts}. Efficient gains in public safety result from replacing  “individualized diagnosis and response” with “aggregate classification systems” used for “surveillance, confinement, and control” \citep{feeley1992new}. Yet the utilitarian, consequentialist rationale for ARAIs worries legal scholars who see their use as conflicting with individual constitutional rights to avoid self-incrimination and receive procedural due process and equal protection \citep{starr2014evidence, starr2015new}. 

In the instrumental pursuit of efficiency, predictive algorithms now carry out the “selective incapacitation” policies popular in the US from the 1980s and 90s \citep{kehl2017algorithms}. Selective incapacitation “selects” or classifies high risk individuals using statistical methods in order to isolate them from the social community \citep{blackmore1983selective}. The method relies on finding highly correlated indices of a behavior of interest, particularly those which reliably discriminate between low, medium, and high risk persons. The ends of public safety are believed to justify the predictive means. \citet{berkbleich2014forecasts} note, “ [if] shoe size is a useful predictor of recidivism, then it can be included as a predictor...Why shoe size matters is immaterial.” The utilitarian believes the net benefits in reduced crime from identifying and incapacitating high-rate offenders outweigh the costs of “wasted imprisonment” on low-rate offenders \citep{mooreDanger1984}. Utilitarians are generally willing to accept more false positives (falsely imprisoning an innocent person) and fewer false negatives (treating a dangerous criminal leniently) to achieve greater public safety. 

In contrast to utilitarian and consequentialist theories of punishment, retributivists look backwards. They argue that if punishment serves to inflict retribution, confer just desert, or express moral condemnation, basing it on prediction is fundamentally misguided. Indeed, \textit{just desert} theories preclude justifying punishment for its instrumental value, such as improving public safety \citep{darley2000incapacitation}. Just punishment is therefore \emph{proportional} to the seriousness of the committed crime, or in the words of philosopher Immanuel Kant, to the “internal wickedness” of the perpetrator \citep{carlsmith2002we}. Retributivists disapprove of algorithmic prediction because it violates the presumption that one must actually do something prohibited by law—not merely be “at risk” of doing something—before losing one’s right to determine one’s own future \citep{feinberg1970doing}. That is, moral blame stems from one’s past choices, not one’s potential future conduct \citep{von1984ethics}. In general, retributivists claim algorithmic approaches conflict with individual rights and moral autonomy, and criticize the narrow focus on predictive accuracy in legal decision-making \citep{underwood1979law, pundik2008evidence, wasserman1991morality}. 

\subsection{Contribution and Scope}
The contribution of this paper is fourfold. First, we collect and organize a diverse literature of ARAI-related issues from machine learning, statistics, sociology, criminology, law, economics, philosophy and related fields. Second, we unite these sources under our newly introduced umbrella concept of \textit{predictive inconsistency}. Predictive inconsistency occurs when an algorithmic system generates disparate predicted scores for the same individual, based on conceptually-justified-but-technically-different choices by ARAI designers. Third, we identify and taxonomize  sources of predictive inconsistency
and relate them to real-world data science and criminal justice choices. Fourth, drawing on extant legal practices and the normative framework of scientific and political pluralism, we propose adding multiverse and specification curve analysis techniques to ARAI development and auditing toolkits in order to estimate a lower bound on predictive inconsistency, as well as to illuminate diverse sources of discretionary bias and quantify their impact on predicted risk scores. 

The paper proceeds as follows. Section \ref{sec:law-ml} considers the nature and value of consistency in various fields and relates legal consistency to our guiding concept of predictive inconsistency. Section 3 describes the ARAI-building process. Section 4 identifies and discusses specific sources of predictive inconsistency related to decisions at the various stages of ARAI development and deployment. Section 5 considers how more complex predictive algorithms combined with new forms of behavioral data may exacerbate predictive inconsistency. To quantify and evaluate predictive inconsistency, Section 6 draws on pluralist scientific and political theories to propose a publicly-justified ``multiverse'' prediction framework for documenting and reproducing forking paths. Section 7 provides our conclusions. 

\section{Consistency in Algorithmic Risk Assessment: Forks versus Knives}\label{sec:law-ml}

Below we describe several notions of consistency and relate them to concepts of justice in order to set the stage for our new term, \textit{predictive inconsistency}. In logic and philosophy, consistency is a property of a set or system of axiomatic propositions, all of which, if true, render the conclusion true \citep{marcus1980moral}. In scientific modeling, a valid model makes predictions consistent with observed data \citep{oreskes1994verification}. In statistics, consistency is a property of an estimator, while in psychometrics, the reliability of a measurement is its consistency across occasions or across items designed to measure the same construct \citep{groves2011survey}. Moving to law, witness testimony and forensic evidence is said to be consistent with a legal theory explaining the facts of the case \citep{kiely2005forensic}. In jurisprudence, consistency denotes coherent judicial decisions \citep{sunsteinRitov2002predictably}, and can refer to consistency of approach or consistency of sentences \citep{lovegrove1997framework}. Consistency of approach signifies that sentences are related to penal aims and case facts in a principled way, while consistency of sentences implies like sentences for like cases. Real-world examples of consistency in US law are captured by principles such as {\it procedural due process} and {\it procedural regularity}, which protect against arbitrary and biased lawmaking targeting specific individuals or groups \citep{kroll2017accountable}.

\subsection{Legal Consistency}
While consistency rests in a dense web of relations to other concepts, we focus on its relation to justice. Early Greek philosophers believed the ideal society, as well as the cosmos, to be ruled by an intelligible, rational and eternal “law and order” best expressed through numerical relations \citep{kline2012mathematics}. Justice involves acting to restore this order or proper balance. Aristotle, for instance, believed the just person develops a disposition for moral discernment through habitual exercise of practical reasoning to particular cases in a variety of contexts \citep{ackrill1988new}.

\rthree{Aristotle's insights are still relevant to judicial reasoning today. Judges must engage in a delicate balancing act that involves inferring judicial principles from individual cases in a consistent way while also maintaining the flexibility to handle new and unexpected circumstances \citep{cane2002responsibility}. Just legal judgments are proportionally balanced in the sense of not being under- or over-fit to the case at hand; this balance provides the desirable property of generalizing to future cases. Yet as Aristotle also insisted, one cannot give \textit{a priori} rules specifying precisely where this balance must lay \citep{ackrill1988new}. As in predictive modeling, the price of rigid consistency with past data (i.e., memorization) is generalizability. Justice requires consistency, but not absolutely:  ``law never requires a judge to sacrifice ‘justice’ on the altar of consistency” \citep[pg. 20]{cane2002responsibility}.} Still, consistent legal decision-making is valuable, even if not absolutely. Consistency may be evidence of the underlying accuracy and objectivity of legal judgments \citep{legomsky2007learning}, as when multiple judges decide similar cases in similar ways. The consistency of law also provides citizens with a degree of predictability to guide them in planning and orienting their lives \citep{hart1961concept}.  This aspect of consistency is related to the concept of the \textit{rule of law} \citep{rosenfeld2000rule} which, in part, makes citizens accountable only to those laws which are publicly promulgated and capable of being followed.

\rthree{Perhaps the most worrying effect of slavish adherence to consistency is its tendency to blind one to novelty and difference, the recognition of which can expand the scope and application of law, as often occurs in landmark cases.} For instance, different judges may examine what appears to be superficially the ``same" type of case and come to contradictory conclusions, yet this disagreement does not always imply judicial inconsistency. A highly experienced judge may have noticed a small but relevant detail---perhaps ``hidden" under the surface---that had escaped the attention of other judges. Once identified, this newly discovered relevant difference \citep{burch2019objectivity} can modify pre-existing rules or principles, or be used to craft new ones, thereby contributing to the evolution and expansion of law and concept of justice \citep{hayek1973law, cane2002responsibility}.

 In practice, however, it is difficult to separate mere difference from inconsistency. Justice means different things to people of different religions and worldviews, and so legal disputes may themselves concern the question of what constitutes a relevant moral or legal difference \citep{rawls2005political}. For this reason, \citet{perezInconsistencyLaw} argues that perfect legal consistency--—a perfectly uniform application of rules--—in a diverse, “pluralistically sensitive” society would result in systemic injustice to certain groups whose moral and political beliefs may conflict with those of legislators and judges. Perfect legal consistency can, for instance, be achieved by repressing freedom of belief and conscience, or by restricting opportunities for good faith dialogue with others. In contrast, some degree of legal inconsistency should be expected and even desired in a pluralistic society, as its existence and eventual resolution hints at the possibility of developing more comprehensive legal principles, moral concepts, and, ultimately, progress in the pursuit of justice.

\subsection{Predictive Inconsistency}
Although we argue some degree of legal inconsistency in pluralistic societies should be expected and even tolerated, inconsistency in science is generally viewed as undesirable. Inconsistency in decision-making results in “noise” and derives from variation over occasions, variation across individuals, or both \citep{kahneman2016noise}. Indeed, the inability to reproduce experimental findings has prompted a methodological ``crisis" in psychology \citep{simmons2011false}. Inconsistent ARAI predictions may invite legal controversy for similar reasons.  Even relatively simple ``actuarial'' risk assessment tools can generate conflicting risk scores for the same individual. For example, in the 2018 case \emph{State v. Gordon}, Gordon was classified in his pre-sentencing report as \textit{high-risk} for sexual re-offense using the three-category SOTIPS tool, while the five-category STATIC-99R classified him as \textit{average risk} \citep{gordonArai}.

Motivated by these methodological and real-world legal concerns, we focus on a new term, \textit{predictive inconsistency.} Predictive inconsistency captures conceptually-justified-but-technically-different choices by ARAI designers leading to disparate predicted risk scores for the same individual. Predictive inconsistency arises when conceptually-similar-but-technically-different data science and data collection choices produce disparate and possibly contradictory predictions. Our notion of predictive inconsistency differs from recent from work in statistics and machine learning focusing on the implications of pipeline underspecification \citep{damour2020underspecification}, the \textit{Rashomon effect} \citep{breiman2001statistical} and various properties of the \textit{Rashomon set} of ``almost equally-accurate predictive models" \citep{dong2020exploring, fisher2019all, semenova2019study}. In line with legal debates around individualized justice, our approach emphasizes the impact of data science decisions on individual predicted scores, not properties of the models themselves. Second, our proposal aims to examine variability across these predicted scores without necessarily requiring actual (ground truth) outcome or labels (except at the model training stage), whereas most work in this area computes predictive accuracy or goodness of fit metrics that assume access to actual outcomes.

Predictive inconsistency is important to understand as it affects public-facing algorithms in high-stakes contexts. Statisticians \citet{gelmanloken2014forkpath} illustrate the problem using Jose Luis Borges’ metaphor of a “garden of forking paths” to describe data analysis decisions and variable human factors they refer to as “researcher degrees of freedom.” These degrees of freedom represent unacknowledged yet non-trivial sources of uncertainty whose impact is not yet well studied \citep{sauerbrei2020state}. Crucially, these data-dependent choices are not laid out in advance, and researchers often do not realize their particular findings are one outcome of many possible forking paths, each representing a different sequence of data analysis decisions \citep{gelmanloken2014forkpath, dwork2015reusable}. Data scientists developing ARAIs face similar problems. Because no general algorithm or inductive bias is guaranteed \textit{a priori} to be the best for a particular problem \citep{wolpert1996lack}, data scientists try a variety of strategies, methods, and algorithms—often with multiple tuning parameters—to find a predictive model delivering the best generalization performance on new, unseen data presumed to come from the same distribution \citep{duwekim2017out}.

We claim the forking paths problem will likely be compounded as the algorithms and sources of data used in ARAI development increase in complexity, and as the variety of actors and institutions involved grows. Just as scientific publications can gloss over analytical missteps and false starts in retrospectively idealized final analyses \citep{pickering2010mangle, latour2013laboratory}, a deployed ARAI embodies an idealized \textit{single path} through what is actually a 
``garden" of many path-dependent and contingent data science practices, such as unacknowledged and undocumented choices by the human factors involved earlier in 
generating, gathering, and cleaning the data, and later in 
developing, testing, and deploying the algorithm. 
Yet judges, corrections officers, and other end-users of the resulting tool's predictions may underestimate the cumulative impact of these discretionary choices and practices on an individual’s predicted risk score. 

We believe the sociotechnical nature of ARAIs and resource limits of real-world criminal justice systems require us to reconsider the value and nature of predictive inconsistency. We should not expect ARAIs to exhibit the level of consistency provided by the controlled laboratory conditions of the natural sciences or the mechanistic precision of the assembly line. A more realistic expectation---aligned with the actual practices of legal reasoning in a diverse society---is that a well-functioning ARAI should reliably assign an individual into a risk category across a variety of ``reasonable" forking paths, which we interpret as representing \emph{good enough} or \emph{reasonable} rather than \emph{optimal} or \emph{rational} descriptions of the same underlying phenomenon viewed from different vantage points. We think this more inclusive approach can bolster the scientific, legal, and political legitimacy of these new and often controversial algorithmic tools.

Our re-framing of predictive inconsistency embraces the democratic spirit of scientific and political pluralism (see e.g.,  \citet{kellert2006scientific, shadish1993critical, kitcher2003science, dewey2012public, rawls2005political, habermas2015between,  bohman2006deliberative}). Broadly speaking, pluralism endorses tolerating some inconsistency and conflict in the transformative search for overlapping consensus on scientific and political issues by deliberating on diverse yet reasonable worldviews, problem representations, classification schemes, analytical methods, economic and social interests and interpretive assumptions. Pluralism challenges the default assumption that ARAIs can or should embody a single, unique logic of decisions leading to a “best” model resulting in a single, clear-cut final prediction (“knives”). Pluralism, in our view, instead engages a diverse team of individuals---in terms of both functional (i.e., cognitive) and identity diversity \citep{hongpagePNASdiversity}---in ARAI design, development and auditing focused on assessing the distribution of predictions generated for an individual from a multiplicity of coequally reasonable models (``forks"). \rthree{Our commitment to democratic pluralism might also warrant the possibility of non-use of an ARAI, particularly when its predictive inconsistency exceeds bounds set by a diverse group of data scientists, domain, and legal experts.}

\enlargethispage{1pc}
\section{Building an ARAI}
\begin{figure}
\centering
\includegraphics[width=1.05\textwidth]{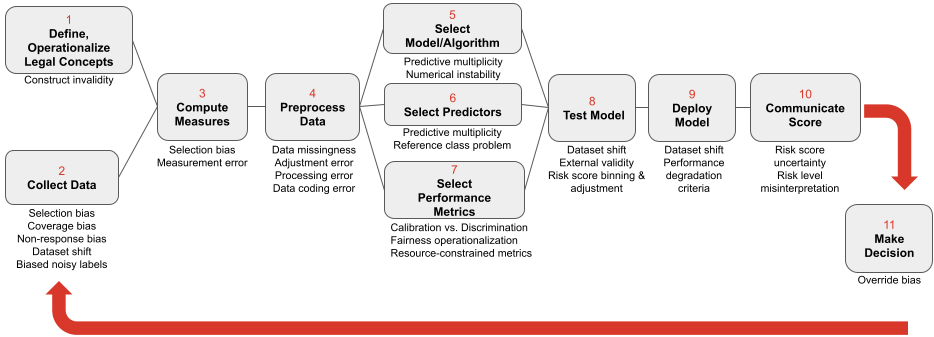}
\caption{\label{fig:pipeline} Steps and potential sources of predictive inconsistency in the development and use of an ARAI. Discretionary ``forking path” decisions by ARAI designers can reduce the tool’s ability to provide consistent predictions for the same individual. Common descriptors for the sources of inconsistency are listed below the step. }
\end{figure}

ARAIs are sociotechnical systems composed of technical artifacts, human agents, and resource-constrained institutions whose rules and norms influence and guide the behavior of human agents \citep{selbst2019fairness, van2020embedding}. We use the general term \emph{ARAI designers} to refer to institutional stakeholders and industry and/or academic software development and data science teams, often with criminal justice domain expertise. To concretize the sociotechnical nature of ARAI design and relate it to data science decisions, we provide a brief description of the ARAI-building process (see Figure \ref{fig:pipeline}). 

First, a data source must be found, representing a sample from a theoretical target population, such as \emph{all prisoners in Minnesota.} These data are often drawn from court or administrative records \citep{ritter2013predicting}. \citet{duwe2014development}, for instance, obtained the data for the MnSTARR tool from the Minnesota Correctional Operations Management System (COMS), based on a sample of 11,375 male offenders released from prison between 2003-2006. Next, ARAI designers must define and operationalize the outcome or construct of interest, such as recidivism—defined broadly as the commission of subsequent crime within a set time period—and decide on a particular time horizon. \citet{duwe2014development} defines recidivism as \emph{reconviction for a criminal offense within four years of release from prison.} Designers also need to decide how many risk categories to include (e.g., low, medium, high) and how much resources they can realistically devote to the different categories. \citet{duwe2014development} uses four risk levels with cutoffs varying by gender and type of recidivism (e.g., sexual or violent). 

Designers then select predictor variables given the data available and theoretical or policy concerns. \citet{duwe2014development} considered over 100 different predictors, keeping just eight using a bootstrap variable selection procedure. Examples include \emph{prior supervision failures}, \emph{total felony convictions}, \emph{drug offense convictions}, and \emph{false information given to police convictions}. ARAI designers typically also consider the relative costs of misclassification (false positives and negatives) with stakeholders. 

After building several models and evaluating their resulting predictive accuracy and risk distributions, designers may adjust aspects of the model or data. Assuming accuracy and risk distributions are appropriate given the resource limitations of the institution developing the tool, the model is then deployed for offenders in the agency’s caseload. For MnSTARR, male inmates are scored at prison intake and once again prior to their release, after updating “dynamic factors” that may have changed during incarceration, such as undergoing chemical dependency treatment \citep{duwe2014development}. This final risk prediction is used to decide the post-release level of community supervision. Lastly, designers monitor the performance over time to ensure its accuracy and reliability.

\section{Potential Sources of Predictive Inconsistency}
This section describes the real-world process of data collection and algorithmic development and illustrates its impact on predictive inconsistency, as shown in Figure \ref{fig:pipeline}. To guide our presentation, we follow a taxonomy of errors affecting data-driven processes commonly used by statisticians and social scientists \citep{groves2011survey}. While textbooks tend to focus on issues of \emph{sampling error} and statistical inference, we emphasize the effects of a variety of \emph{non-sampling errors} such as \emph{construct invalidity}, \emph{measurement error}, \emph{processing}
and \emph{data coding error}, \emph{coverage bias}, \emph{non-response bias}, and \emph{adjustment error} on predictive inconsistency. 
  Note that discretionary choices in earlier steps can propagate and exacerbate predictive inconsistencies in later steps. An example is measurement error at the data collection stage that later impacts predictors and response variables, thereby affecting a model’s predictive performance \citep{kuhn2013applied}.

\subsection{Defining and Operationalizing Constructs Into Empirical Measures}
The law ``teems with devices that defeat uniformity and  predictability''\citep{easterbrook1992abstraction}. Due to its vast scope of application, law possesses an “open texture” focused on defining general classes of persons or acts \citep{hart1961concept}. That is, for practical reasons, law cannot in advance define every conceivable criminal act in society and therefore considers only general classes of acts (e.g., murder, battery, fraud) under which particular acts may fall.  Legal systems, like most natural systems, are open, not closed, and so legal assertions require an element of intuitive judgment and interpretation that both precludes absolute proof or demonstration and resists explicit automation \citep{schauer2009Lawyerthink}. In contrast, statistical algorithms must presume the validity of a particular interpretation or application of a general label (i.e., categorical outcome variable) to make inferences and predictions at all \citep{barocasmlbook2019fairness, selbstbarocas2018intuitive}. But whether the resulting predictions rightfully apply to anything in the “real world” depends on the degree of correspondence between internal model elements and external objects of interest \citep{selbst2019fairness}. Crucially, the validity of such correspondence can never be absolutely verified \citep{oreskes1994verification}.

The often implicit act of interpretation in ARAI development requires explicitly operationalizing theoretical and unobservable legal, social, and moral concepts known as \textit{constructs} \citep{groves2011survey}. \rthree{\textit{Operationalizing} a construct involves specifying a corresponding set of uniquely specifiable physical or mathematical operations \citep{bridgman1927logic}.} Yet assessing the validity of measurements of essentially and unobservable constructs is not straightforward \citep{borsboom2004concept}. For that reason, a modern and pragmatic view holds that \textit{construct validity} is a holistic, evidential judgment of how well a measure captures its intended construct and supports the interpretation of a score, including action taken on the basis of this interpretation \citep{messick1995validity}. Generally, the higher the stakes of assessment, the more important construct validity becomes \citep{downing2003validity}. One implication is that ARAI designers cannot evade the ethical issues of operationalizing “essentially contested" constructs and values such as \emph{fairness} \citep{mittelstadt2019principles, wallachjacobs2021measurement, friedler2021possibility}.

\begin{table}
\caption{Various criteria used to measure recidivism. Definitions obtained 
from \citet{breitenbach2010creating, blumstein1971problems, maxfield2005measuring,jones1997recidivism, BureauOfJusticeStats2002, angwin_2016}.
\label{tab:reciddefinition}}
\centering
\fbox{%
\begin{tabular}{l|l}
\hline
\textbf{Criteria}      & \textbf{Options used by different studies/systems}          \\ \hline
Events             & Arrest, Conviction, Incarceration     \\ \hline
Degree             & Felony, Misdemeanor, Public Ordinance \\ \hline
Time Periods       & 2 Years, 3 Years, 5 Years         \\ \hline
Since                  & Previous Crime, Arrest, Incarceration, Conviction           \\ \hline
Inclusion Criteria & In-state/Out-of-state                 \\ \hline
Predicted Outcome Type & Time-to-recidivate, Probability of Recidivism, Hazard Ratio \\ \hline
\end{tabular}
}
\end{table}

\rthree{The operationalization of recidivism risk is one such contested and complex process that impacts predictive inconsistency.} No generally accepted legal definition of recidivism exists, and the literature devoted to discussing, defining, and approximating various operationalizations of recidivism has a long history \citep{rector1958factors}. We note, however, that existing definitions of recidivism share three features. Each definition has a starting event from which the measurement of recidivism commences, e.g., release from prison. Second, each definition has a measure of failure following the starting event, e.g., a subsequent arrest. Third, each definition has a window of recidivism, that is, a follow-up period within which the offender’s behavior is observed \citep{zgoba2015recidivism}. Table \ref{tab:reciddefinition} illustrates the diversity of definitions.

According to the US National Institute of Justice, recidivism refers to ``a person's relapse into criminal behavior, often after the person receives sanctions or undergoes intervention for a previous crime.” Some operationalize it as the duration between two events, e.g., days from release date to the point of the first warrant date \citep{breitenbach2010creating}. Others measure it by a dichotomous “reconvicted/not” or “new arrest/not” within a certain time period from some event \citep{jones1997recidivism, maxfield2005measuring}. A Bureau of Justice Statistics report uses four measures of recidivism: \emph{rearrest}, \emph{reconviction}, \emph{resentence to prison}, and \emph{return to prison with or without a new sentence within a three-year period following the prisoners’ release}, and further distinguishes between “in-state” and “out-of-state” recidivism \citep{BureauOfJusticeStats2002}. The much-publicized ProPublica study of bias in risk assessment tools defined it as a \emph{new arrest within two years of the original crime for which the subject was assessed, while discounting any minor offenses and municipal ordinance violations} \citep{angwin_2016}. This definition is problematic because it counts subjects who were arrested but were not convicted, and those whose charges were dropped. The choice of recidivism measure also leads to different selected models and algorithms: predicting the expected time until recidivating calls for a different type of model than for the probability of recidivating in the next five years. The same model cannot produce both. 

Different choices can lead to different predictive performance and therefore to inconsistent predicted risk scores. Changing the time horizon in the definition of recidivism changes the \emph{base rate} of the phenomenon (i.e. the underlying proportion in the population of interest), which in turn affects relevant predictive performance measures. For example, \citet{rice1995violent} show changes in the definition of “violent recidivism” to include any new violent crimes within a horizon of 3.5, 6, and 10 years (resulting in base rates of 15\%, 31\%, and 43\%, respectively), causes various performance measures of the Violent Risk Appraisal Guide (VRAG) model to fluctuate dramatically. Nevertheless, end users of an ARAI may not be aware that different definitions might have led to different risk scores for the same individual. 

\subsection{Collecting Data and Computing Measures}
Statistics and machine learning textbooks often assume no errors in the outcome labels and thus narrowly focus on issues of sampling error \citep{hand2006classifier}. While increasing the sample size can reduce sampling error, many sources of inconsistency mentioned in the ARAI literature are actually instances of \emph{non-sampling error} \citep{lohr2011sampling}. Common non-sampling errors that can arise in the data collection step are \emph{selection bias}, \emph{measurement error}, \emph{nonresponse bias}, and \emph{coverage bias}. Data collection and preprocessing procedures are major sources of non-sampling error and are hard to identify and thus control. Merely having “big data” does not reduce non-sampling error. 

The bias and noise added due to non-sampling error can propagate through to later stages of algorithm development \citep{suresh2019framework}. Examples of non-sampling error are confusing survey questions, unobserved social and economic pressures influencing persons to respond in systematically different ways, and optical character recognition (OCR) devices generating systematic errors in the transcription of handwritten text. 

\emph{\bf Coverage bias} arises when building predictive models of well-defined target populations based on court and administrative records. Coverage bias relates to mismatches between the \emph{sampled population} and \emph{target population} \citep{groves2011survey}, or in ML terms, the algorithmic \emph{development population} and \emph{use population} \citep{suresh2019framework}. Depending on what is available, some ARAIs will rely on data from convicted populations, while others on arrested populations \citep{ritter2013predicting, duwe2014development}. Yet, in the UK, only about 2\% of all crimes ever face sentencing decisions \citep{ashworthsentence2005}. And some are arrested without having committed a crime. 

Further, poverty and racial disparities in arrests may be reflected in the administrative data used to train risk prediction algorithms, a phenomenon known as “biased noisy labels” \citep{fogliatonoisy2020}. Rich, well-connected criminals with access to expensive legal representation may be arrested but convicted less often. Computer crimes, identity theft, tax fraud, and a variety of other “low-priority” white collar crimes are notoriously difficult to detect and so may be under-represented if only arrests or convictions are counted \citep{cole2007american}. Ideally, data should be representative of attributes (e.g., “propensity to recidivate”) of the population or construct of interest. To the extent power and wealth disparities, or policing strategies, are reflected in the data, measures based on such data will lead to issues of \emph{construct invalidity} and \emph{measurement error} and actions taken on the basis of the predicted scores may invite critical scrutiny. 

\emph{\bf Selection bias} is a concern when relying on non-randomly selected samples to estimate patterns in the population \citep{heckmanselectbias1979}. In ML, the related term \emph{\bf{dataset shift}} describes the situation where the joint distribution of inputs and output differs between the training and test datasets \citep{quineroDatashift2009}. Datasets used to train and test predictive algorithms for use in criminal law can suffer from over-representation of some populations and under-representation of others \citep{zavrvsnik2021algorithmic}. Such misrepresentation leads to inconsistencies when deployed to new members of under-represented populations, or even to completely different populations. An example is when predictive models trained on mostly male inmates are applied to female inmates \citep{hannah2001taking}. Modern ARAIs, however, increasingly create separate predictive models for men and women \citep{duwe2014development}. 
Yet systems developed and trained for one target population are sometimes used in other geographical, demographic, temporal, and decision-making contexts, such as an algorithm developed for decision on prison releases being applied for probation decisions \citep{kehl2017algorithms}. \citet{berk2019machine} lists many other causes of dataset shift including \emph{parole board personnel turnover}, \emph{judges losing elections}, \emph{changes in the number of police and prisoners}, and \emph{changes in drug markets}, among other causal factors. As an example of potential generalization issues, the Public Safety Assessment tool used pre-trial training data from 300 US jurisdictions, but was applied statewide in Kentucky, Arizona, New Jersey, and Utah, where pre-trial populations of ethnic subgroups are likely different from the overall population of jurisdictions \citep{stanfordPolicy2019b}. An example of good methodological practice comes from \citet{tollenaar2019optimizing}, who trained algorithms on Dutch conviction data and evaluated their external validity on a different target population consisting of North Carolina prisoners.

\enlargethispage{2pc}
\emph{\bf Measurement error}  is the difference between a measurement and the true value of a quantity \citep{crowder2020introduction}. It is typically described using \emph{random}, \emph{systematic} or \emph{differential error} models \citep{luijken2019impact}. In statistical inference, measurement error can lead to model nonidentifiability, blurring the meaningful interpretation of model parameters \citep{grace2016statistical}. But measurement error can also amplify predictive inconsistency \citep{frenay2013classification}, as real-world data gathering in criminal justice settings involves noisy measurement, affecting data labels used during ARAI training and later at the time of scoring \citep{duwerocque2017effects}. In plea bargains, for instance, a person is arrested for one crime but charged for a less serious one \citep{breitenbach2010creating}. Despite presumed innocence, poorer persons without high-quality legal representation may be more likely to accept plea bargains \citep{ashblakeinnocent1996}. Due to constraints on legal resources, the practice of prosecutorial discretion resembles ``satisficing" \citep{albonetti1986criminality} and can result in reduced police booking charges (e.g., reducing a felony to a misdemeanor); these reductions may also vary by race and familiarity with the informal workings of the criminal justice system \citep{albonetti1990race, albonetti1992charge}.

Financial incentives can contribute to measurement error by influencing the behavior of human data collectors. Organizational pay-for-performance schemes may motivate police to incorrectly classify crimes and describe arrests. Gaming techniques used by police to “hit their targets” include “choosing not to believe complainants,” “recording multiple incidents in the same area as a single crime,” and “downgrading incidents to less serious crimes” \citep[pg. 128]{muller2019tyranny}. \citet{maltzFBI1999} mentions the FBI hierarchy rule to avoid double counting. If two types of crime occur in the same incident, only the category of the most serious crime is counted: e.g., “If a convenience store robbery results in the death of the store clerk, this would be classified as a homicide rather than a robbery—because homicide is a more serious crime than robbery” \citep{maltzFBI1999}. Such gaming strategies lead algorithms to learn predictive relationships more reflective of record-keeping limitations and wishful thinking than of reality.

Feedback loops in the criminal justice system \citep{ensign2018runaway} not only make data collection and causal interpretation of data complex, but they can also affect measurement error. Due to variation in the amount and quality of data collected,  some individual or group-level records are harder to predict than others \citep{rudin2022interpretable}, resulting in greater predictive inconsistency. One explanation for this phenomenon touches on larger structural justice issues, as some minoritized subgroups come into contact with administrative agencies more often than others, leading to class imbalances in resulting datasets \citep{buolamwini2018gender, dignazio2020datafem}. If risk predictions are combined with arrest data generated by predictive policing algorithms, the resulting increased police surveillance can lead to more arrests and recorded criminal incidents although the underlying crime rate has not changed (e.g., \citet{na2013police}). When such data are then fed back into judicial decision making algorithms, they create a self-sustaining feedback loop that reflects more the nature of the crime sampling process than actual patterns of criminal behavior \citep{partnershipAI2019}. Judges, parole and probation officers may also treat those predicted “high risk” in systematically different ways \citep{koepke2018danger}, leading to \emph{differential measurement error}. For example, algorithm predictions in medical settings can lead doctors to undertake different forms of measurement (e.g., self-report of body weight versus physician-directed measurement on a scale) \citep{luijken2019impact}. 

Differential measurement effects can be related to an individual's race or gender \citep{mullainathanober2017, suresh2019framework}, raising social justice concerns and resulting in predictors of unequal predictive power for some subgroups compared to others.  For instance, the predictor \textit{criminal history} might be more predictive of recidivism for older defendants than for younger ones without a detailed criminal history, or for those in living or working areas with a history of intense police surveillance \citep{browne2015dark}. Worried by the explosive growth of mass incarceration in America, sociologists have analyzed US administrative, survey, and census data between 1969-1999 and concluded that incarceration rates for young, low-skilled black men are so high as to resemble a rite of passage into adulthood, on par with college graduation or military service \citep{pettit2004mass}. Criminology research also suggests that incarceration may actually slightly increase one’s probability to recidivate \citep{cullen2011prisons, durlauf2011imprisonment}. Science and technology studies scholars point to these and other examples as evidence for the role of ARAIs in contributing to a \textit{matrix of domination} \citep{collins2002black} that reproduces pre-existing social, historical, and economic inequities \citep{benjamin2019race, dignazio2020datafem}. 

Pre-parole questionnaires used as input to ARAIs are also susceptible to measurement error. They are subject to all the standard sources of measurement error in surveys, including how they are worded, what time and in what setting they are taken, who administers them, and various faulty recall effects \citep{whittle2018measurement}. As an example, parolees in the Pennsylvania Corrections Department are asked simple yes/no questions regarding their past behaviors, such as whether they have ever had a drug or alcohol problem \citep{barryjester2015}. But they are not asked to indicate the severity of the problem or what constitutes a “drug or alcohol problem.” Even more, self-reports can be intentionally manipulated by respondents themselves or by data collectors. COMPAS thus embeds a “data validity check” in self-report questionnaires to identify respondents suspected of “lying, sabotage, or incoherent responses” \citep{brennandieterich2018}. Given the strong incentive offenders have in avoiding a “high risk” classification, ARAI designers should periodically assess algorithms and proxy measures for their susceptibility to strategic manipulation, particularly differential gaming strategies by subpopulations. If detected, designers may add randomness or update the algorithm more frequently \citep{bambauer2018algorithm}. Yet if proxy measures are chosen wisely, gaming behavior can be made to incentivize the socially-beneficial self-improvement of offenders \citep{kleinberg2020classifiers}.

\subsection{Data Preprocessing}
Data scientists may increase predictive inconsistency by employing various pre-processing strategies to improve model performance, especially when trying to predict rare events. In social science and psychometrics, these kinds of issues broadly fall under \emph{adjustment error}, \emph{processing error} and \emph{data coding}, including handling missing values \citep{groves2011survey}. Although we do not discuss transforming or standardizing predictors, these common operations can also introduce predictive inconsistency \citep{sauerbrei2020state}. 

\emph{\bf Data missingness} involves missing values or incomplete data. When such missingness is systematic (e.g. refusal to respond to sensitive questions about criminal behavior), it may lead to the exclusion of specific populations. Data scientists must then make decisions about whether and how to impute missing values, keep only complete observations, or do nothing. Each decision may have different and unpredictable effects on the final model.  For example, approximately 10\% of randomly-selected inmates declined to participate in the data collection efforts for the COMPAS Reentry risk assessment tool \citep{breitenbach2010creating}. Censored data (unobserved outcomes) is a frequent problem in criminal justice settings because those whose outcomes are observed may be systematically different from those whose are not \citep{berk2019machine}. Missing data can cause further inconsistency at prediction time, if the to-be-predicted record has missing predictor values. Machine learning solutions, such as training multiple models with different subsets of predictors \citep{saar2007handling}, are another source of inconsistency.

A frequent pattern in ARAI data is imbalanced outcome variables. For example, in a pretrial risk tool examined by \citet{eckhouse2019layers}, only 3.8\% were arrested for a violent crime and 4.9\% failed to appear in court. When the outcome variable is imbalanced, a common preprocessing step is to group together types of rare categories to achieve better balance among the classes. But such grouping comes with a cost. By replacing examples of drug-related or domestic violence with a crude dichotomous measure, such as “violent” felony crimes, important criminogenic distinctions are obscured and risk predictions may not properly reflect the context of assessment, leading to inconsistent predictive performance \citep{breitenbach2010creating}.  

If the outcome variable is imbalanced, predictive algorithms can fail to learn to distinguish between the rare and majority classes. Popular preprocessing strategies to deal with this problem include oversampling or undersampling of rare or overrepresented classes (e.g., women in COMPAS Reentry), using cost-sensitive learning, or synthetically generating new records of the rare events \citep{weiss2013foundations}. \textit{Discrimination aware} pre-processing can also involve re-labeling or resampling/reweighting methods \citep{d2017conscientious}. A simpler method limits the modeled population to a relevant, less-imbalanced subpopulation that contains the class of interest \citep{weiss2013foundations}, such as those committing violent crimes. A meta-analysis of 68 studies of violence ARAIs found this strategy improved predictive accuracy \citep{singh2011comparative}. This approach also finds legal support in \citet{mayson2017dangerous}, who argues pre-trial detention may only be appropriate for persons predicted high risk of violent recidivism, not merely any type of recidivism. But doing this requires precise criteria for deciding which subdomains count as “sufficiently interesting.” Different data scientists may come to different conclusions, thus adding a source of inconsistency in the resulting predictions.

\subsection{Variable Selection and the Reference Class Problem}\label{RefClassVarSelect}
\emph{Predictive multiplicity} occurs when competing models produce conflicting predictions \citep{marx2020predictive}. One source of such multiplicity (i.e., predictive inconsistency) arises when selecting variables assumed to be predictive of the outcome. Generalizable inductive inference requires developing robust predictive models that trade off model complexity with model fit (the so-called \textit{bias-variance tradeoff}) \citep{hastieTibsElements2009}. An overly-complex model with many predictors can be made to fit arbitrarily well to a given dataset, but in doing so loses the ability to predict well for new, unseen data. To combat this, a key data science strategy is the process of variable selection \citep{hastieTibsElements2009}, where predictors based on specified criteria are included or excluded in the model. Examples include lasso and stepwise regression, pruned classification trees, and the bootstrap selection procedure used to build the MnSTARR ARAI. Yet small changes in variable inclusion criteria affect which predictors go into the model, thus acting as a source of predictive inconsistency \citep{heinzevarselect2018}. 

Variation in variable selection procedures impacts predictive inconsistency in individualized sentencing because different “evidentiary estimates” could be given to the same person, each on seemingly justifiable grounds \citep{rhee2007probability}. Predictors included and excluded in the final model determine the criteria potentially defining one’s reference group and thus one’s predicted risk. The more predictors selected, the narrower the potential scope of an individual’s reference group used to calculate risk. But the more specific the reference class, the fewer the individuals belonging to the class, thus increasing uncertainty of the risk estimates. To illustrate, we can calculate risk based on gender alone, or on a larger set of predictors such as gender, age, and zip code. In the former, the reference class is other people of the same gender, whereas in the latter the reference class is others of the same gender, age, and in the same zip code. 

\citet{cheng2009practical} argues that variable selection mirrors the \textit{reference class problem}, which occurs when we “want to assign a probability to a single proposition, X, which may be classified in various ways, yet its probability can change depending on how it is classified” \citep{hajek2007reference}. The 1995 case of \emph{United States v. Shonubi} provides a vivid example of its implication for legal risk assessment, though in a slightly different context from standard ARAIs. Tasked with estimating the unknown amount of heroin Mr. Shonubi—a Nigerian national and US resident—smuggled into JFK Airport on seven previously undetected trips, \citet{cheng2009practical} says, “The court could have considered the amount carried by all drug smugglers at JFK, all Nigerian smugglers regardless of airport, or smugglers in general.” Each reference class assignment results in different estimates. The accuracy of these estimates was important because they decided which sentencing guidelines would be applied to his case, potentially adding several extra months or years in prison. The ensuing legal battles incited a lively discussion of when statistical evidence is admissible as “specific evidence” to be used in an individual’s sentencing proceedings \citep{tillers2005if}. 

Prior data collection choices determine the granularity of possible reference classes and the number of possible subsets of predictors evaluated during variable selection, but ideally variable selection should not depend on the particular data at hand \citep{heinzevarselect2018}. Currently, however, different jurisdictions have access to administrative data of varying granularity and quality. For instance, is \emph{nationality} recorded when drug incidents are reported? If not, then it cannot be used as a predictor. Deciding an individual’s reference class is not only a philosophical issue highlighting the indeterminacy \citep{leiter2007objectivity} of \emph{a priori} defining legally relevant differences justifying differential treatment, but also impacts predictive inconsistency. 

\subsection{Selecting Performance Metrics}
\subsubsection{Discrimination vs.~Calibration}
AUC (Area Under the (ROC) Curve) is a popular performance measure used in ARAIs, yet misunderstanding what it measures can contribute to predictive inconsistency. As \citep{hand2009measuring} notes, “choosing a measure which does not reflect [one’s aims] could lead to incorrect conclusions.” AUC estimates the probability that a random positive observation (e.g. recidivator) ranks higher than a random negative (non-recidivator) \citep{flach2019performance}. Algorithms with higher AUC scores have better predictive validity: \citet{duwerocque2017effects} define an “adequate” level for ARAIs as 0.70 or higher. But AUC is appropriate only when one’s predictive goal is an accurate ranked ordering of risk scores, otherwise known as \emph{discrimination} \citep{singh2013predictive}. AUC says nothing about the accuracy of the predicted score. \emph{Calibration}, in contrast, focuses on the estimation of an exact probability (and compares it to the percentage observed in the test set) \citep{tollenaar2019optimizing}. As \citet{fawcett2006introduction} explains, an algorithm with high AUC “need not produce accurate, calibrated probability estimates; it need only produce relatively accurate scores that serve to discriminate [between] positive and negative instances.” The same model may perform better at one or the other task, depending on the nature of the training data and structure of measurement error \citep{whittle2018measurement, luijken2019impact}. Lastly, AUC comparisons can be misleading when the underlying ROC curves cross, which commonly occurs when comparing multiple classifiers \citep{hand2009measuring}.  

\enlargethispage{1pc}
\subsubsection{Formalizing Fairness Criteria}
Problems of fairness in ML have been formulated as constrained optimization problems \citep{corbett2017algorithmic, zemel2013learning}, which require trading-off predictive accuracy and various formal notions of fairness \citep{berkroth2021fairnessml}. Fairness metrics can focus on either individual or group outcomes and may be used at pre-processing, in-processing, or post-processing stages \citep{berk2019machine}. Our discussion mostly centers on post-processing metrics and does not consider those based on ``causal" comparisons of actual with counterfactual outcome distributions \citep{kusner2017counterfactual}. We also leave open philosophical questions of whether it is possible or desirable to formalize fairness or justice \citep{derrida1992force, greenhu2018myth}.

Algorithms can have disparities in predictive accuracy for various racial or gender subgroups, raising questions of bias and unfair discrimination in the data collection process itself \citep{angwin_2016}. For instance, one measure of discrimination compares true positive rates across groups, a fairness metric known as \textit{equal opportunity} \citep{hardt2016equality}. But merely observing different rates of predictive accuracy among subgroups is not necessarily evidence of “unfair” algorithms. Against initial reports of racial discrimination in the COMPAS ARAI \citep{angwin_2016}, the makers of the tool, Northpointe Inc., argue that when base rates of recidivism differ significantly among racial groups, symmetry in the error rates across groups cannot be achieved \citep{dieterich2016compas}. This debate highlights whether unequal base rates are better understood as symptoms or causes of unjust social conditions, conditions in which some social groups are systematically and unjustifiably more likely to be subjected to ARAIs than others. Likewise, such issues cast doubt on the appropriateness of narrowly treating fairness as a property of algorithms, independent of larger social contexts \citep{selbst2019fairness}.

 At the same time, fairness metrics may help identify deeper structural problems at the data collection and operationalization stages. \citet{eubanks2018automating} illustrates this situation using a Pennsylvania-based tool used to predict cases of child abuse. Children in poor families tend to come into contact with public services such as child protective services, Medicaid, and drug and alcohol treatment programs more often than those from wealthy families, resulting in more extensive data collection for poor families, in terms of both the number of observations and the number of features. In such situations, the ARAI is apt to confuse “parenting while poor with poor parenting” \citep[pg. 127]{eubanks2018automating}. The counter-intuitive implication is that ARAIs trained using sparse data from wealthier families may exhibit greater predictive inconsistency, prompting ARAI critics to call for discontinuing their use or delegitimizing their predictions. Meanwhile, ARAI proponents could point to the low predictive inconsistency for poor families as a reason to continue using such tools. Yet it seems unfair to subject some families---namely poorer ones---to more algorithmic decision-making and state surveillance and intervention because more and better quality data has been collected on them. In other words, the mere fact of having more data should not count as a \textit{relevant difference} justifying more surveillance. 

As one might expect, basing formal predictive measures on contested legal and ethical concepts is difficult, if not impossible. Imposing fairness criteria can itself impede the social goal of minimizing expected violent crime because satisfying the criteria can only be achieved by releasing more high-risk defendants \citep{corbett2017algorithmic}. There is thus an inherent tension between treating individuals equally and achieving “algorithmic fairness” by setting race-specific decision thresholds. \citet{berk2019machine} also points out that by not including “discriminatory” information such as race or gender in the algorithm (i.e., “fairness through unawareness”)—or even their proxies, such as neighborhood—the reduced predictive power may result in greater numbers of dangerous persons released back into communities. \citet{kleinberg2016inherent} and \citet{chouldechova2017fair} demonstrate that in normal situations, no single algorithm can achieve the desired fairness properties of \emph{calibration}, \emph{balance for the negative class}, and \emph{balance for the positive class} simultaneously. Only in extremely rare cases of “perfect risk assignment” by the algorithm and equal base rates among subgroups can a single algorithm satisfy the above-mentioned definitions of fairness. As \citet{dieterich2016compas} remark, this is unlikely to hold in practice. 

The discussion of competing fairness metrics also highlights a deeper issue: the distinction between individual and group justice \citep{binns2020apparent, mitchell2021algorithmic}. Realistically, one must choose the least unfair of a set of unfair algorithms \citep{speicher2018unified}. For example, \citet{dwork2012fairness} base their operationalization of fairness on an Aristotelian, individualized notion of “nearest neighbor parity,” capturing whether persons represented as similar in predictor space receive similar predictions. Yet many of the approaches detailed above rely on the resulting group-level parity of calibration or false positive rates. Satisfying both forms of fairness requires tradeoffs: minimizing between-group unfairness can increase within-group unfairness \citep{speicher2018unified}. Consequently, “hard choices” among fairness metrics cannot be made independently of the complex ethical considerations of individual versus group justice \citep{binns2018polyfairness}. Equally-justifiable metrics can lead to disparate predicted scores. ARAI designers may thus consider generating predicted scores for a variety of fairness definitions and operationalizations based on plurality of legal and philosophical assumptions.

\subsubsection{Lift: Resource Constrained Ranking}
Although ARAIs are often justified by claims of cost-saving and efficiency in resource allocation \citep{duwe2014development, ritter2013predicting}, surprisingly few predictive applications use lift to evaluate classifier performance \citep{shmueli2019lift}. Lift is the ratio of the true positive rate in the \textit{top-n} sample (where \textit{n} is subject to budget or resource constraints) to the true positive rate in the entire test set. The greater the lift, the better the classifier performs compared to random targeting.  

In criminal justice contexts, public agencies often have budget and resource constraints limiting their ability to act on every prediction made, a fact with implications for performance evaluation and predictive inconsistency. An ARAI chosen for its superior performance as judged by \emph{capacity-unconstrained} metrics, such as AUC, may be judged inferior when lift is evaluated. As noted above, lift measures a classifier’s ranking (i.e. discriminative) ability, just like AUC. But unlike AUC, lift captures the classifier’s ability to maximally “skim the cream” from a subset of individuals \citep[pg. 136]{shmueli2017data}. The amount of resources that can be invested into acting on the predicted scores decides the size of the subset. As an illustration, \citet{duwekim2017out} note that random forests and logistic regression performance appear very similar in terms of AUC in predicting recidivism, yet random forests much more accurately rank the highest risk offenders. 

\enlargethispage{4pc}
In our view, it makes little sense to rely on capacity-unconstrained metrics, such as AUC, in constrained scenarios in which real criminal justice systems operate. Because lift explicitly considers budgetary limits it is well-suited for judicial and policing contexts. But lift requires greater coordination between data scientists and legal domain experts, as they would need information about budgetary constraints and operating scenarios. Lift also raises questions of justice and fairness: is it reasonable to rank people and act only on the top-ranked? When predicted probabilities are—in an absolute sense—low, acting on a predefined top x\% of cases could violate intuitive notions of fairness.

\subsection{Adjusting and Communicating Risk Scores}
ARAI designers convert risk probabilities into risk levels for decision-makers to interpret and act on. Although judges typically have discretion in choosing not to follow ARAI recommendations, evidence from case studies suggests they often prefer to have them if available \citep{hartmann2021uncertainty}. An ARAI might display output for an individual similar to the following: “Risk of Recidivism: Medium; Failure to Appear: Low; Risk of Violence: High” and provide a recommended supervision level \citep{angwin_2016}. ARAIs generally, however, give no indication of the uncertainty associated with predicted risk levels \citep{partnershipAI2019}. 

Despite little evidence that categorical risk levels promote better informed decision making \citep{scurich2018case}, it is common practice to bin continuous probabilities into discrete risk levels in order to ``aid practitioner interpretation" \citep{chiappa2018causal}. In some US states, such as Kentucky, practitioners are required to create risk groups \citep{stanfordPolicy2019a}. Nevertheless, there may still be a large, unaccounted for gap between a predicted probability and categorical score. Depending on relative misclassification costs, an algorithm developer might decide to classify an individual as “high risk” if the predicted probability ranges between 0.50-0.99. \citet{mayson2017dangerous} suggests that binning and threshold decisions involving preventative restraint should ideally be based on comparisons with the general population of non-defendants. We note that binning decisions are related to construct validity and measurement error \citep{hanson2017communicating}.

There are pros and cons to creating risk groups \citep{hilton2015communicating}. Discretizing risk levels discards potentially useful information, but can increase consistency across judges who might interpret raw probabilities in different ways. \citet{gastwirth1992statistical} cites a study of judges in the Eastern District of New York revealing varying interpretations of probability assigned to important legal standards of proof, such as “preponderance of evidence” and “[evidence] beyond a reasonable doubt.” On the other hand, a more precise and finer breakdown of risk, such as displaying percentile ranks, absolute recidivism rates, or risk ratios \citep{hanson2017communicating}, may reduce the consistency of judicial decision-making processes. But access to these more fine-grained distinctions could theoretically promote more individualized sentencing. 
In either case, when predicted probabilities are converted into risk levels, this choice is usually neither driven by legal or theoretical considerations nor aligned with the outcome measure used in training, which is typically an arrest or conviction, not a “risk level.” In fact, agencies often adjust algorithms’ predictions ex-post until predicted risk scores are broadly aligned with the desired or expected distribution of risk in the target population \citep{ritter2013predicting}, which is only vaguely estimable. An ARAI designed to help allocate probation supervision levels where 90\% of individuals receive a “high risk” classification is useless in practice. But these ad-hoc adjustments and risk groupings create inconsistency across systems and applications. 

Finally, as hinted at earlier, the decision to train an algorithm for discrimination or calibration can result in different predictive models being declared most predictively accurate. This is because small differences in relative ranks (e.g., 1-2) may hide large absolute differences in probability (e.g., 0.99-0.01). In other words, the best model for a calibration task may not be the best model for discrimination task: it will depend on what one is trying to predict. The absolute risks of some crimes may be so low (e.g., a 4.9\% failure to appear rate) that converting predicted probabilities into relative ranks or “high, medium, and low risk-levels” is misleading to end-users \citep{eckhouse2019layers}. 

\section{Possible Futures for Algorithmic Risk Assessment}
\label{sec:futures}

Despite methodological debate about the “exceptional variation” of individual predicted scores and whether this imprecision warrants their disuse \citep{cooke2010limitations, imrey2015commentary}, current ARAIs are mostly limited to relatively interpretable and computationally stable statistical regression models. The next generation, however, will likely rely on more automated and flexible ML algorithms with built-in data ingestion and variable selection capabilities \citep{duwerocque2017effects, slobogin2017principles}, such as those used in predictive policing. We also expect these models to use more granular, micro-level behavioral data on individuals and key criminal justice actors and decision makers, such as police and judges. As noted earlier, judges have discretion in choosing to follow predicted risk scores, resulting in a “gap” between algorithmic recommendation and the final judicial decision. \citet{green2020false} details research showing how some judges in Cook County, Illinois are systematically biased in favor of increased detention, and how defendants’ race is correlated with decisions to increase individuals’ predicted crime risk. We refer to this phenomenon as \emph{override bias} (see Figure \ref{fig:pipeline}, step 11). \rthree{Further, various institutional and economic pressures---not to mention other psychological factors such as anchoring bias \citep{chang2016anchoring} and ARAI output misinterpretation \citep{yacoby2022if}---may affect how and when judges diverge from the algorithm's recommendation. To better understand the relevant factors involved and promote greater accountability,} \citet{koepke2018danger} suggest requiring judges to justify these divergences and that agencies begin collecting data on when and how they occur.

If a source of data improves the ability to reliably discriminate among risk levels, for-profit companies and under-funded public agencies have clear incentives to use them to more efficiently allocate resources \citep{berk2019machine}. A shift is already underway in some risk-modeling industries such as credit scoring and car insurance, where auto insurance policy rates can be augmented with behavioral driving data and GPS locations. A recent study by \citet{ayuso2019improving} used telematic data to design auto insurance policies, emphasizing the need to find “new variables of risk exposure and driver behaviour.” \citet[pg. 172]{berk2019machine} refers to these new individually weak, but collectively powerful sources of predictive information as “dark structure,” arguing “there is substantial structure in criminal behavior that current thinking in criminal justice circles does not consider.” But if things move in this direction, the hidden sources of predictive inconsistency we mentioned will persist and perhaps become even more difficult to detect.

In terms of ML algorithms, classification and regression trees are a likely next step due to their relative transparency and interpretability. \citet{berkbleich2014forecasts} suggest random forests as a candidate ML technique, which also offer predictor importance scores and the ability to easily employ asymmetric misclassification costs in model training \citep{berk2019machine}. \citet{kleinberg2018human} use gradient boosted trees to predict Failure to Appear (FTA) cases in New York City. Random forests are used by both the Harm Assessment Risk Tool (HART) in the UK \citep{oswald2018algorithm} and a COMPAS tool used by some US prisons for new inmate classification \citep{brennan2009evaluating}. We note, however, that ML algorithms, even tree-based algorithms, can be  numerically unstable: re-running the algorithm with the same data but a different random initialization seed can give a different output \citep{berk2019machine}. Even using different software can lead to a different result, thus increasing predictive inconsistency. 

Paralleling the evolution in predictive algorithms, the next generation of ARAIs will rely on new and more finely-grained sources of behavioral data as input \citep{werth2019risk}. As digital era governance and IT infrastructures expand, we foresee a move from using only well-understood input measures (e.g., criminogenic factors), towards “features” derived from fusing multiple data sources (e.g. social media), with data in multiple formats (e.g., numerical, text, image, video, network, etc.). In China, using such behavioral big data is already common in law enforcement and tightening up the government’s social and political control \citep{jiang2018chinese}. In the US, California has trialed GPS monitoring of high-risk parolees, and in some areas DUI probationers are given remote monitoring anklets \citep{ridgeway2013pitfalls}. In Taiwan, \citet{you2018sobermotion} designed a portable breathalyzer device to reduce drunk driving by logging and analyzing complex behavioral data and reporting them to a local probation office. These sources of behavioral data can easily be adapted and included as predictors in future ARAIs. But as the sources and variety of input data grow, so too do worries of robustness under conditions of dataset shift \citep{berk2019machine}. 	

New and more complex sources of data may also conflict with contemporary understandings of the rule of law in Western democracies \citep{liu2019beyond}. The distinction between using algorithms to explain or predict \citep{shmueli2010explain} raises legal questions about the relevance and defensibility, accountability, and transparency of high-stakes blackbox predictions when employed by private companies and public authorities \citep{oswald2018algorithm}. For instance, algorithm audits in the healthcare domain have uncovered evidence of racial bias due to spurious correlations stemming from construct invalidity \citep{obermeyer2019dissecting}.  Unless citizens know they are subjected to such algorithms and they have some way of challenging the results (which itself requires some degree of explainability), predictive algorithms pose the risk of reproducing, not resolving, societal inequities. 

\section{Towards Illuminating and Quantifying Forking Paths}\label{sec:quantify}

As ARAIs use more sophisticated algorithms and harvest more complex sources of behavioral data, how can we illuminate and estimate predictive inconsistency in ways compatible with and justified by legal, political, and scientific practices? We suggest a pragmatic compromise based on a philosophy of pluralism. Pluralist ARAI design tolerates some level of predictive inconsistency by viewing a well-functioning ARAI as reliably assigning an individual into a risk category across a variety of ``reasonable" forking paths. This approach aims to cancel out systematic biases of individuals or groups of data scientists by combining predictions made across a variety of agreed-upon reasonable forking paths constituting the design, decision, representational, and attentional biases of data scientists from diverse walks of life. \rthree{Inspired by the ``notice and comment'' rulemaking procedures of US government agencies \citep{mulligan2019procurement}, the democratic pluralism we endorse insists that deliberation on the``reasonableness" \citep{binns2018publicreason} of a path  be responsive to public criticism from those likely to be subjected to predictions (or their chosen appointees). In this way, individuals can influence decisions affecting their interests \citep{nagel2013representation}, or at least be heard in a due process. The predictions generated through the democratic and technologically-aided procedure we propose may better achieve the normative legitimacy characteristic of ``procedurally fair" legal judgments \citep{habermas2015between, tyler2006people}. We thus advocate for ARAI design based on publicly-contestable, multi-path ``forks" rather than monological, single-path ``knives." }

But not all forking paths are under the control of ARAI designers. For example, data scientists have little to no control over the quality of the administrative data, or how corrections officers conduct interviews and assessments. Further, judges are not required to follow the ARAI’s recommendation and may even misinterpret its output. Local legal requirements also constrain construct operationalization. Yet many paths are under the designers' control. We therefore distinguish between forking paths that are quantifiable by ARAI designers and those that are not. Quantifiable paths comprise a variety of generic data science decisions, which can vary according to local legal and dataset-driven considerations. These paths can be used to help illuminate the various sources and impact of predictive inconsistency on individual risk prediction scores. 

\begin{table}
\label{tab:fpaths}
\caption{Sample template for deliberating on the reasonableness of quantifiable “forking path” choices by ARAI designers, prior to plotting individual-level predictive inconsistency with specification curves. }

\begin{tabular}{{|p{2.8cm}|p{4cm}|p{5.2cm}|}}
\hline
\textbf{Step in ARAI Development} &
  \textbf{Data Science Choices (examples)} &
  \textbf{Evaluating Reasonableness of Path} \\ \hline
Defining and operationalizing constructs &
  \begin{tabular}[t]{@{}l@{}}Event type\\ Time period\\ Geographic area\end{tabular} &
  Check local law and domain knowledge (e.g., criminology)\\ \hline
Collecting data and computing measures &
Data scientists generally have less control over data collection and quality & 
  \begin{tabular}[t]{@{}l@{}}Use or create data sheet \\ Check for:\\ Coverage bias\\ Measurement error\\ Selection bias\\ \end{tabular} \\ \hline
Data preprocessing &
  \begin{tabular}[t]{@{}l@{}} Data imputation \\ Over-/under sampling \\ Group rare categories\\ Bin continuous features\\ Remove outliers\\ Transform predictors \end{tabular} &
Use model cards and exploratory data analysis. Investigate reasons for missingness and/or class imbalance \\ \hline
Model/variable selection &
  \begin{tabular}[t]{@{}l@{}}Regression model \\ Random forest \\ Stepwise procedures\\ Bootstrap \\ Regularization\\ Add pairwise interactions\end{tabular} &
  Examine interpretability and numerical stability (e.g.,~different random seeds and train/test splits) \\ \hline
 Selecting performance metrics  &
  \begin{tabular}[t]{@{}l@{}}AUC\\ Brier scores\\ Lift\\ Fairness metrics\end{tabular} &
  \begin{tabular}[t]{@{}l@{}} Determine:\\  Calibration or discrimination goal?\\ Institutional budget constraints?\\ Individual or group fairness?\end{tabular} \\ \hline
Adjusting risk scores and communication  &
  Create discrete risk bins &
Check local law for binning guidance and budgetary limits \\ \hline
\end{tabular}
\end{table}

The first step is to identify and document the multiplicity of paths involved in the development of an ARAI, as outlined in Section 4. We then envision turning identified forking paths into self-contained, shareable, searchable and reproducible computational environments using tools and workflows adapted from software engineering \citep{forde2018reproducible, ragan2018binder}. Model-level ``predictive multiplicity” metrics \citep{marx2020predictive} can be combined with auditing tools such as \textit{data sheets} \citep{gebru2021datasheets} and \textit{model cards} \citep{mitchell2019model} to help document data and modelling choices relevant to a specific prediction. The goal of this step is to facilitate internal and third-party ARAI auditing and possibly even foster civic engagement by creating publicly accessible ARAI registries, as several major cities have done \citep{venturebeatAIRegistry2020}. 

\rthree{After identifying and documenting candidate paths, our pluralist approach calls for a diverse group of domain experts to deliberate on which paths are ``reasonable,'' drawing on their combined technical, legal, and theoretical knowledge. Again, any agreed upon definitions, ranges, or thresholds should be responsive to and contestable from those likely to be subjected to predictions, or their chosen representatives (i.e., a public data science advocacy group). Table 2 provides a basic template for evaluating the reasonableness of a currently-deployed ARAI or one in development. Assuming consensus has been reached on a set of reasonable forking paths,} and these paths have been identified and documented in reproducible environments, the next step is generating predictions for individuals. We propose adapting the methods of \emph{multiverse analysis} \citep{steegen2016increasing} and \emph{specification curve analysis} \citep{simonsohn2020specification}, a technique giving a high-level, visual overview of multiverse analyses. Multiverse analysis stems from applications in psychology \citep{simmons2011false} and the analysis of experimental fMRI data, where up to 35,000 different forking paths may be at play \citep{carp2012plurality}. 

Although multiverse analyses generally focus on parameter estimation and statistical inference, we instead suggest adapting these techniques to individual-level predictions for a specific ARAI under development or audit. The essential idea is to evaluate subject-level predictive inconsistency by generating reasonable forking paths and obtaining the set of resulting prediction scores for each subject. The score distribution for an individual is then plotted using specification curves. During ARAI development and auditing, individual-level score distributions can be used for estimating overall predictive inconsistency stemming from manipulable forking paths and for researching sources that might be eliminated. Further, the score distributions can also be used at the time of decision making to, for instance, illuminate person-specific predictive inconsistency levels and help end-users determine the appropriate level of credibility to attach to the ARAI’s score. \rthree{Similar to the suggestion of \citet{marx2020predictive}, when an ARAI exhibits an unreasonable degree of predictive inconsistency, we might choose not to make predictions for that person, or implement the model at all.}

To reiterate, because of our focus on individual-level prediction scores, the key modifications to multiverse analysis include (1) creating a predictive inconsistency holdout set, ideally based on a purposive or representative sample of the population of interest located where the ARAI is likely to be deployed; (2) determining ``reasonable" minimal predictive and fairness performance metric thresholds, based on deliberation among data scientists, domain, and legal experts (e.g., whether AUC, Brier scores or lift should be used and their acceptable thresholds or ranges; which fairness metric is used, etc.); (3) adapting specification curves to display subject-level predicted scores. Specification curves helpfully reveal both model specification details (i.e., which forking paths went into an individual's prediction) and corresponding predicted scores. We envision the results of these analyses open to public comment and scrutiny.

Multiverse analysis can also help to align ARAI design and auditing with principles and practices of legal responsibility aimed at preventing and repairing harms in society. In civil law, for instance, domain specific, community-based standards of ``reasonable conduct" and care are used to assess responsibility for harmful outcomes \citep{cane2002responsibility}. When evaluating negligence claims in domains such as medicine, a common test for establishing causation entails asking whether, ``but for the negligence of the defendant, the plaintiff would not have been injured" \citep{epstein1973theory}. A further consideration is whether abnormal or deviant conditions were present \citep{hart1985causation}. An \emph{ex post} multiverse analysis can thus help reveal that an individual’s predicted risk scores would likely have fallen into some range, “but for” a particular data science decision. With the help of a diverse community of domain experts to specify “reasonable” forking paths, multiverse analysis can encourage socially accountable ARAI design by providing standards to assess the reasonableness of a particular design choice. Besides providing useful “contrastive” or counterfactual causal explanations \citep{miller2019explanation} of predicted risk scores, multiverse analysis can guide efforts to improve data collection and quality and indicate construct invalidity when irrelevant factors have an unexpected and disproportionate impact on an individual’s predicted risk score.

Prediction-oriented multiverse analysis is limited by applying only to quantifiable forking paths amenable to manipulation by data scientists. Analyses may therefore only provide a lower bound for an ARAI’s predictive inconsistency for a given individual. Privacy concerns, strategic gaming considerations, and issues of intellectual property law may also restrict the feasibility and generality of this collaborative approach. Citizens and ARAI end-users might also lack the requisite data science and legal knowledge or motivation to participate in developing community-based standards of reasonableness. \rthree{For this reason, a specialized government agency could be tasked with setting such standards (see, e.g.,  \cite{scherer2015regulating}). In any case, we encourage interdisciplinary research into not only the sources of predictive inconsistency affecting ARAIs and similar sociotechnical systems, but also into the larger legal and institutional decision-making processes in which they are embedded.}

\section{Conclusion}
\rthree{Although controversial}, ARAIs increasingly influence decision-making in criminal justice systems worldwide. Predictive inconsistency captures the conceptually-justified-but-technically-different decisions by ARAI designers that reduce the ability to provide consistent predictions for the same individual. Key sources of predictive inconsistency include construct invalidity, measurement error, processing and data coding error, coverage bias, non-response bias, and adjustment error. These errors relate to the way ARAI designers define, operationalize, collect and preprocess data, select predictor variables, choose models and performance metrics, test and deploy them, and communicate risk scores in many and often conflicting ways. Grounded in the normative framework of scientific and political pluralism, we propose multiverse and specification curve analyses as steps towards reproducible, explainable and democratically accountable ARAI development and auditing. These methods reveal how reasonable deviations (forks) from a single development path (knives) affect the variability of a particular prediction. They also provide new means for contesting predictions and advancing ARAIs' scientific, political and legal legitimacy.  

Criminal justice systems operate with limited resources, and ARAIs promise more efficient and consistent decisions. Yet these practical benefits must be weighed against the potential harms of unjust treatment stemming from algorithmic blindness to  morally relevant differences in individual persons and cases. Questions of whether and how ARAIs can be used in a just, responsible, transparent and innovative manner will not be resolved by more sophisticated technology, more rigorous mathematical formalism, or more abstract ethical principles alone. Tolerating some degree of predictive inconsistency may be the price we must pay to live in a democratic, dynamic, and pluralistic society where legal and social practices, moral concepts, technologies, policing and carceral strategies and even the definitions of crimes evolve over time and \rthree{are subject to the changing needs and interests of citizens. Whether the resulting imprecision negates the ostensible social and legal utility of ARAIs we alone cannot say. This is largely a political issue  \citep{wong2020democratizing} that requires not only an appeal to technical expertise, but to normative authority as well.} Ultimately, however, we support adjusting the tools and techniques of predictive modeling to fit the complex and plastic nature of persons, law and society, rather than adjusting persons, law and society to fit the idealized and rigid nature of predictive modeling.

\section*{Acknowledgements} 
We thank Mark Shope for his valuable comments and feedback and the Associate Editor and three reviewers for their many thought-provoking suggestions. Shmueli and Greene were partially funded by Taiwan Ministry of Science and Technology [Grant 108-2410-H-007-091-MY3]. Lin was partially funded by Taiwan Ministry of Science and Technology [Grant 111-2628-H-007-001].

\bibliographystyle{rss}
\bibliography{ARAI}
\end{document}